\documentclass[sigconf]{acmart}
\renewcommand\footnotetextcopyrightpermission[1]{}
\setcopyright{none}
\acmConference[WWW '21]{}{April 19-23, 2021}{Ljubljana, Slovenia}

\usepackage[linesnumbered,ruled,lined]{algorithm2e}
\usepackage{algcompatible}

\usepackage{multirow}
\usepackage{subfig}

\AtBeginDocument{%
  \providecommand\BibTeX{{%
    \normalfont B\kern-0.5em{\scshape i\kern-0.25em b}\kern-0.8em\TeX}}}

\setcopyright{acmcopyright}
\copyrightyear{2018}
\acmYear{2018}
\acmDOI{10.1145/1122445.1122456}

\acmConference[Woodstock '18]{Woodstock '18: ACM Symposium on Neural
  Gaze Detection}{June 03--05, 2018}{Woodstock, NY}
\acmBooktitle{Woodstock '18: ACM Symposium on Neural Gaze Detection,
  June 03--05, 2018, Woodstock, NY}
\acmPrice{15.00}
\acmISBN{978-1-4503-XXXX-X/18/06}

\begin{document}

\title{Graph Neural Network Based VC Investment Success Prediction}




\author{Shiwei Lyu}
\email{lvshw@shanghaitech.edu.cn}
\affiliation{ShanghaiTech University}

\author{Shuai Ling}
\email{lingshuai@shanghaitech.edu.cn}
\affiliation{ShanghaiTech University}

\author{Kaihao Guo}
\email{guokh@shanghaitech.edu.cn}
\affiliation{ShanghaiTech University}

\author{Haipeng Zhang}
\email{zhanghp@shanghaitech.edu.cn}
\affiliation{ShanghaiTech University}

\author{Kunpeng Zhang}
\email{kpzhang@umd.edu}
\affiliation{University of Maryland}

\author{Suting Hong}
\email{hongst@shanghaitech.edu.cn}
\affiliation{ShanghaiTech University}

\author{Qing Ke}
\email{qke100@syr.edu}
\affiliation{Syracuse University}

\author{Jinjie Gu}
\email{jinjie.gujj@antfin.com}
\affiliation{Ant Financial Services Group}


\begin{abstract}
Predicting the start-ups that will eventually succeed is essentially important for the venture capital business and worldwide policy makers, especially at an early stage such that rewards can possibly be exponential.

Though various empirical studies and data-driven modeling work have been done, the predictive power of the complex networks of stakeholders including venture capital investors, start-ups, and start-ups' managing members has not been thoroughly explored. We design an incremental representation learning mechanism and a sequential learning model, utilizing the network structure together with the rich attributes of the nodes. In general, our method achieves the state-of-the-art prediction performance on a comprehensive dataset of global venture capital investments and surpasses human investors by large margins. Specifically, it excels at predicting the outcomes for start-ups in industries such as healthcare and IT. Meanwhile, we shed light on impacts on start-up success from observable factors including gender, education, and networking, which can be of value for practitioners as well as policy makers when they screen ventures of high growth potentials.
\end{abstract}

\ccsdesc[500]{Information systems~Data mining}


\keywords{venture capital investments, start-up success prediction, graph neural network, incremental graph representation learning}


\maketitle

\section{Introduction}

Predicting success of start-ups carries important implication for a wide variety of stakeholders, ranging from industry practitioners such as entrepreneurs and venture capital (VC) firms, to government officials. Specifically, VC firms provide financing to start-ups that have exceptional growth potential, bearing above-average risks in exchange for above-average returns, whereas governments design policy instruments to boost growth of entrepreneurial firms, including innovation grant programs and government-managed VC funds. Many highly successful tech companies, including Amazon, Google, and Facebook, all received funding from venture capitals before they became major players in the marketplace and their early-stage investors who bore the largest risks reaped exponential returns. Exiting through an acquisition or an IPO is the main source of profit for VC firms, with IPOs yielding the most revenue~\cite{gompers2001venture, sorensen2007smart}. For example, as an early investor of Google, Sequoia Capital received over \$4 billion in value after selling their stake in Google in 2015, representing about 320 times return from their initial investment of \$12.5 million in 1999. Previous studies suggest returns of VC-backed investments present a skewed distribution. Kerr \textit{et al.} show that about 60 percent of VC-backed investments between 1985 and 2009 terminated at a loss and 10 percent of them generated a return more than five times the capital invested~\cite{kerr2014entrepreneurship}. Furthermore, existing evidence suggests poor performance of many policy tools targeted at entrepreneurship growth. Grilli \textit{et al}. find government-managed VC does not have positive impacts on success of start-ups~\cite{grilli2014government}. Given the high variance of VC investment returns, it helps with commercial value creation as well as with improvement of policy efficacy if one can identify start-ups with great potentials, especially at a very early stage when they are fledgling or even just untested prot\'eg\'es. Statistics in our data show that 63.44\% of the start-ups receive second-round funding and among which only 13.54\% get acquired and 1.88\%  go up for IPOs.

Many factors that may lead to successful VC investments have been explored, associated with the start-ups~\cite{bohm2017business, li2001does, begley2001socio}, the founding teams~\cite{unger2011human, baum2004picking, franke2006you}, the VC firms~\cite{sorensen2007smart, tian2012role}, the investors~\cite{huang2015managing}, and the VC investment market~\cite{nanda2013investment}. These studies mostly look into correlation between characteristics of VC firms or start-ups and success of the venture without much focus on predicting success of start-ups. The availability of large-scale detailed VC investment data at an individual level from sources including Crunchbase\footnote{\url{http://www.crunchbase.com}} and Pitchbook\footnote{\url{http://www.pitchbook.com}} has provide an opportunity of data-driven approaches to screening projects~\cite{zhong2019venture} and predicting their acquisitions~\cite{xiang2012supervised} or next rounds of funding~\cite{sharchilev2018web}.

\vspace{-1pt}

Prior works have explored various factors that might have impact on start-up success predictions. In particular, a perspective of networks formed by connections among stakeholders has recently gained great attention. For instance, 
Hochberg \textit{et al.} construct a firm-level VC network, with the nodes being the VC firms and the edges representing the co-investment relationships – if two VC firms invest in a same start-up, an edge is formed between the two firms~\cite{HOCHBERG-whom-2007}. They discover that better-connected VC firms are more likely to succeed with regard to investment outcomes. 
There has also been work on identifying potentially successful VC firms with a PageRank-like metric in a bipartite network of VC firms and start-ups with edges representing investment relationships~\cite{gupta2015identifying}. 
Bonaventura \textit{et al.} model the labor-flow among start-ups -- members of a start-up may be former members of other start-ups, and find nodal centralities of start-ups have predictive power of their economic performances~\cite{Bonaventura-pred-2020}. 
These approaches that rely on analytical expertise to extract features can hardly exploit all potentially helpful structural information. Furthermore, these studies have not touched the node attributes (e.g. a person's education background and a start-up's industry) which are yet to be fully integrated with the network structural information to make better predictions.

To better understand the relationship among nodes in a network, many graph embedding techniques have been proposed where each node is represented as a low-dimensional vector while the local and global structural information is well preserved~\cite{tang2015line}. Especially with the advancement of Graph Neural Networks (GNN) on tasks of link prediction or node classification, we have witnessed many successful implementations and their downstream applications in the area of computer vision, natural language processing, healthcare, and many others~\cite{qi20173d,beck2018graph,choi2017gram,fout2017protein}. Kipf and Welling~\cite{kipf2016semi} propose a CNN-like graph neural network to capture structural information in graph data; Veli\v{c}kovi\'c \textit{et al.}~\cite{velivckovic2017graph} introduce graph attention networks (GATs) that use attention mechanisms to aggregate information from neighbors. Researchers also model graph evolution by first constructing several networks in a sequential manner and learning representations for each network separately. Note the subsequent network is a super set of previous networks. Then a time-series-like model is built to learn dynamic behavior of nodes~\cite{sankar2020dysat}. This method can be inefficient especially when the network grows larger, as it involves much unnecessary computing, where newly added nodes are not likely to affect representations of the distant nodes.

In this study, we propose an incremental graph representation learning to predict whether an early-stage start-up would eventually succeed (IPO or acquired). We learn node-level representation in each time period using self-attention and optimize these representations through fine-tuning via supervised link prediction and node classification. We also model sequential dependencies of node representations across time periods. Finally, these learned representations are used for success prediction, together with attributes of nodes in the VC investment network. Overall, our method makes the following main contributions:
\begin{itemize}
    \item We are among the first studying the essential problem of VC investment outcome prediction with a graph neural network that incorporates both node attribute and the network structural information. Experiments conducted on real-world dataset demonstrate that our method outperforms state-of-the-art baselines, up to 1.94 times the performance of real investors.

    \item We develop and implement an incremental updating mechanism for efficient graph neural network representation learning with supervised fine-tuning, which improves the performance as compared to other unsupervised and/or static representation learning. 

    \item We identify factors associated with positive VC investment success predictions given by our model
    that can be of interest from the perspective of financial analysis and predictive analytics. Female investors and managing members of start-ups, though far outnumbered by their male counterparts, demonstrate a positive influence. Other factors include investors and managing members having higher education level and being well-connected in the network, which is consistent the conclusions from previous studies~\cite{HOCHBERG-whom-2007,ewens2020early}.
\end{itemize}
\vspace{-3pt}
For the rest of the paper, we review related works that inspire this study in Section~\ref{sec:related owrk}. We then formally define the start-up success prediction problem in Section~\ref{sec:problem definition} and describe our proposed model in details in Section~\ref{sec:Method}. After that, we present the experimental results and compare against various baseline models in Section~\ref{sec:exp}. Finally, we discuss our findings in Section~\ref{sec:discussion} and conclude the paper in Section~\ref{sec:conclusion}.
\vspace{-4pt}
\vspace{-4pt}
\section{Related Work}
\label{sec:related owrk}
Our work is inspired by three threads of research: VC investment, complex network analysis and graph convolutional neural networks.

\vspace{-8pt}

\subsection{VC investment}
\label{sec:vc investment}
We first examine how VC investment success is defined in the literature.

\vspace{-6pt}

\subsubsection{VC investment success}
\label{success_definition}
Different studies define VC investment success differently. A loose standard consists of positive outcomes within a time window: (i) receiving a subsequent round of funding~\cite{sharchilev2018web}; (ii) acquired by other firms~\cite{xiang2012supervised}; (iii) or going IPO, in which the investors can sell their equity and possibly make a profit. From (i) to (iii), the difficulty and the chance of profitability increase. While (i) does not guarantee profitability, (ii) and (iii) are the main sources of profit~\cite{gompers2001venture, sorensen2007smart}. Around a half of VC-backed start-ups receive a subsequent round of funding, while only around 15\% get acquired or go IPO according to our statistics. In other words, a standard consisting of (ii) and (iii) is more strict and challenging. Combinations of these outcomes in given periods of time form the basis of most success definitions in the literature.
\vspace{-6pt}
\subsubsection{Success factors}
A variety of factors that are associated with success of VC-backed start-ups have been examined by existing research in finance, such as the business models~\cite{bohm2017business}, product innovativeness~\cite{li2001does}, and locations of the start-ups~\cite{begley2001socio}, personality~\cite{unger2011human}, prior start-up experience~\cite{Gruber-look-2008, Gompers-perf-2010}, and professional and educational backgrounds~\cite{baum2004picking, franke2006you} of the founding teams, the syndication~\cite{tian2012role} and value-adding expertise~\cite{sorensen2007smart} of the VC firms, the intuition of the investors~\cite{huang2015managing}, and the environment of the VC market~\cite{nanda2013investment}. These studies are mostly focused on establishing causalities and seldom touch prediction tasks.


\vspace{-4pt}

\subsubsection{Prediction}
Recently, large and detailed datasets from providers such as Crunchbase and Pitchbook make machine learning techniques applicable for relevant predictions. Features about the investors, start-ups, and their founding teams as well as the Web mention data is utilized to predict whether a start-up with early investment would secure a further round of funding within a year~\cite{sharchilev2018web} or whether a start-up would be acquired~\cite{xiang2012supervised}. Comparing with predicting a next round of funding in the near future, predicting the long-term outcome of a particular start-up address more uncertainty. It is worth noting that to simulate the real-world scenario, the selection of training and testing data is crucial but sometimes overlooked. For instance, Sharchilev \textit{et al.} fix a date and make a prediction at that date for each candidate start-up, which lowers the difficulty as a start-up near its next round of funding would inevitably gain more attention from the Internet~\cite{sharchilev2018web} while the 10-fold cross-validation used by Xiang \textit{et al.} leaks future data into training set~\cite{xiang2012supervised}.


Another thread of work aims at easing the burden of project screening by recommending start-up projects to investors. Instead of predicting whether a start-up would succeed, researchers try to predict what projects a VC firm would invest in from a recommender system perspective~\cite{zhong2019venture, zhang2015link}. However, whether a start-up would succeed remains as an unaddressed question that can yield high rewards for investors.
\subsection{Complex network analysis}
\label{sec: complex network}

Different VC firms or investors can invest in a same start-up and managing members of a start-up may have previously worked for different companies. Such relationships create networks that evolve along time and some have been shown to be instrumental in predicting the success of VC investments. 
Hochberg \textit{et al.} investigate the co-investment network of VC firms, with nodes being the VC firms and the edges representing the co-investment relationships between VC firms, finding that better-networked VC firms, measured by centrality metrics, are more likely to exit successfully~\cite{HOCHBERG-whom-2007}. 
In a similar fashion, Gupta \textit{et al.} construct a bipartite network of VC firms and start-ups and investment relationships and propose a PageRank-like metric to identify successful VC firms over time~\cite{gupta2015identifying}.
From the start-up's perspective, researchers form a network of start-ups linked by employee flows and start-ups with high closeness centrality are more likely to receive positive outcomes~\cite{Bonaventura-pred-2020}. They argue that the flow of employees is associated with transfer of knowledge across start-ups. 
Whereas most of these studies compute predefined network metrics, richer structural and temporal information may be missed out among these engineered features.


\subsection{Graph Neural Networks}

In recent years, there is an increasing interest in extending deep learning approaches to handle graph data and they could be a natural fit for predicting on the VC networks that connect investors, start-ups, and their members. Specifically, graph embedding techniques are widely developed to represent individual nodes in a low-dimensional space while capturing complex network structural information. In our context of success prediction, we model it as a binary node classification problem where Graph Convolutional Neural Network (GCN) has been proved to be very effective in achieving good performance.


\vspace{-4pt}
\subsubsection{Graph embedding}

Prior graph representation learning studies usually adopt a similar idea of skip-gram to learn node embedding that maximizes the likelihood of co-occurrence in fixed-length random walks~\cite{tang2015line,perozzi2014deepwalk}. Recently, several neural network architectures have been proposed with great success in graph embedding~\cite{hamilton2017inductive,velivckovic2017graph,kipf2016semi,sankar2020dysat}. For example, convolutional neural networks~\cite{kipf2016semi} along with attention mechanism has been tailored to aggregate information from neighbors for graph learning~\cite{velivckovic2017graph}. Such graph evolution operations might be useful in the our success prediction of a start-up, as indicated by~\cite{Bonaventura-pred-2020}. On the other hand, 
incrgraph learning can indeed improve many downstream graph related applications. For example, Sankar \textit{et al.}'s implemented a new neural architecture with structural and temporal self-attention for node representation learning, which improves their link prediction tasks~\cite{sankar2020dysat}. However, it requires a node's past link evolution in order to get the node's embedding which is not feasible in the scenario that predictions are made only for brand new nodes (early-stage start-ups).
\vspace{-4pt}
\subsubsection{Node classification}
Traditional node classification is typically formed as a supervised learning task where various features such as node attributes, network structural information of a node (e.g., centralities), and group-level characteristics of a node (e.g., communities)~\cite{tang2016node,bhagat2011node} are manually extracted. Recent development in graph neural networks greatly facilitate this task by modeling node attributes and network structures at the same time. Further, graph attention networks (GATs)~\cite{velivckovic2017graph} is proposed to treat nodes in different weights with an attention mechanism. It has been successfully applied for node classification tasks, achieving the state-of-the-art performance with a certain degree of interpretability. We adopt their architectures for our task in this study.

\section{Problem Definition}
\label{problem_definition}
\label{sec:problem definition}
We now formally define the success prediction problem of a start-up's VC investment. 

\begin{figure}[h!]
\centering
\includegraphics[width=6cm]{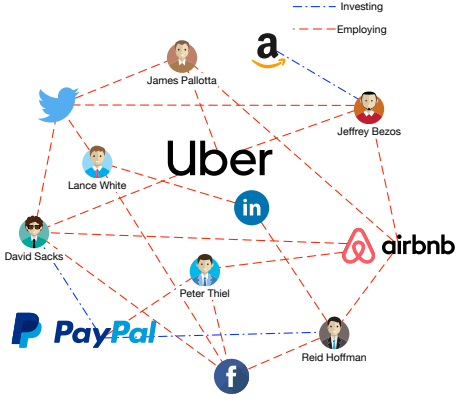}
\caption{An illustration example of a VC investment network, consisting of persons (investors and managing members of start-ups) and start-ups. The edges represent investing or employing relationships.}
\label{fig:Network exapmle}
\vspace{-10pt}

\end{figure}

Let $\mathcal{G}=\{G^0 ,G^1, G^2, \cdots, G^T\}$ denote a sequence of VC investment bipartite graphs over $T$ time periods (e.g., 24 months in this study). In each graph $G^t = (V^t, E^t)$, $V^t$ is the set of nodes added in the $t^{th}$ time period, a node $v\in V^t$ represents either a start-up or a person (an investor or a managing member of a start-up); $E^t$ is the set of edges added in the time period $t$ and $e\in E^t$ represents an investing or employing relationship between two nodes (see the illustration example in Fig.~\ref{fig:Network exapmle}). $G^0$ includes all nodes and edges up to our starting point. $G^t$ ($t>0$) is an incremental update for the $t^{th}$ period. The typical updates are addition of new nodes and new edges. We do not consider deletion and changes which are rarely seen in our context.


Formally, we define our prediction task as follows. Given $\mathcal{G}^t = \{G^0 ,G^1, G^2, \cdots, G^t\}$ and node attribute data (e.g., investor demographics, start-up industry), we predict whether a start-up that receives the first-round of VC investment at the $t^{th}$ period will succeed. Here, our definition of success is that the start-up either is acquired or undergoes IPO within 5 years since its first-round of funding~\cite{HOCHBERG-whom-2007}.
\vspace{-10pt}

\section{Method}
\label{sec:Method}

\begin{figure*}[h!]
\centering
\setlength{\abovecaptionskip}{4pt}
\setlength{\belowcaptionskip}{-10pt}
\includegraphics[width=17cm]{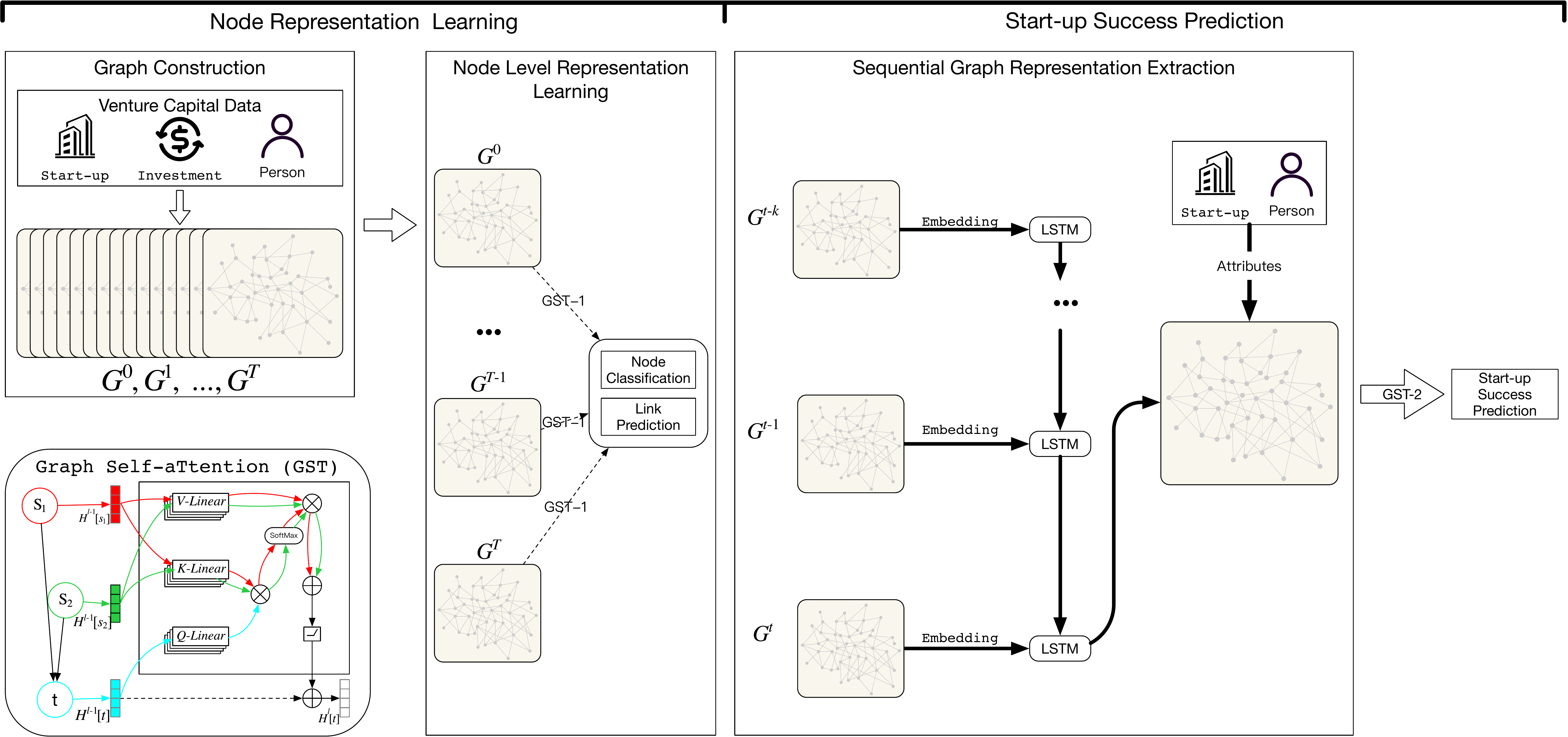}
\caption{The overview of our proposed method.}
\label{fig:framework}
\end{figure*}

In this section, we describe the details of our proposed method, which considers both temporal (incremental graph representation learning) and structural information (graph representation learning with attention) to make success prediction of a start-up's VC investment.  As illustrated in Figure~\ref{fig:framework}, our method consists of five major components: graph construction, graph self-attention learning, incremental representation learning with fine-tuning, sequential graph representation learning, and success prediction.

\begin{itemize}
\item Construction of a VC investment network: we build a sequence of VC networks where each is a bipartite graph representing different relations among persons and start-ups. The initial network at time 0 $G^0$ is all investing activities up to a certain (pre-specified) time period. The rest networks are incrementally constructed. Please see details in Section~\ref{sec:network-construction}.

\item Graph self-attention: we leverage self-attention mechanism to implement a node-level representation learning for an individual VC investment network. See details in Section~\ref{sec:GSAT}.

\item Incremental graph representation learning (GST): we design an incremental updating method to learn the node representation of VC investment networks which are incrementally changed. We leverage a link prediction and node classification-based algorithm for supervision to optimize and fine-tune the learned representation from GST. See details in Section~\ref{sec:increment-updating}.

\item Sequential representation modeling: to capture the temporal relationships among node representations across time periods, we implement a simple recurrent neural network (e.g., LSTM) to better understand their historical dependencies for improving subsequent success prediction. Please see details in Section~\ref{sec:sequential-modeling}.

\item Success prediction: A simple binary classifier is implemented to make the final success forecast, while incorporating node-level attribute data along with the learned representations from previous steps. See details in Section~\ref{sec:prediction}. 

\end{itemize}

\subsection{Construction of VC investment network}
\label{sec:network-construction}
Similar to the network constructed by Bonaventura \textit{et al.}~\cite{Bonaventura-pred-2020}, our VC investment network is a bipartite graph, consisting of two types of nodes (i.e. persons or start-ups). The links can only occur between persons and start-ups, representing the relationships of investing or employing. Figure~\ref{fig:Network exapmle} is an illustration example of this network. In our study, we first use all investment-level activities up to a certain time point (e.g. $t_0$) to build an initial network, denoted $G^0$. We apply network representation learning to obtain initial low-dimensional embeddings for all nodes in $G^0$. This network grows dynamically as more investments occur in subsequent time periods. Embeddings of existing nodes will be accordingly updated when network structure changes (e.g., new nodes and links added).

\subsection{Graph self-attention}
\label{sec:GSAT}
In this section, we introduce the graph self-attention approach for node-level representation learning. Basically, it aggregates information from neighbors to generate new representation of nodes, which is similar to many previously established methods, such as GCN~\cite{kipf2016semi} and GATs~\cite{velivckovic2017graph}. GCN treats nodes' neighbors equally, while GATs applies the attention mechanism to capture different impacts on neighbors. Along this thread, self-attention mechanism has been proposed to enhance the vanilla attention mechanism in many downstream NLP applications~\cite{vaswani2017attention}. This also inspires us to design a Graph Self-aTtention (GST) neural network with self-attention to aggregate information from investors and managing members to obtain better representations for start-ups in the VC investment network.

Specifically, GST works as follows. Assume all nodes' embeddings at the $(l-1)$-th iteration are already learned, denoted $H^{(l-1)}[*] \in \mathbb{R}^{d}$. For each edge ($s - t$), we calculate the importance of source node $s$ to its target node $t$ (a.k.a. attention). Here we adopt an $h$-head attention mechanism, where the value of each attention head is computed separately. The $i$-th head attention $\text{ATT-head}^{i}(s,t)$ is formulated as:
\begin{equation}
\nonumber
    \begin{aligned}
    K^{l}(s) &=W_K^{l}\left(H^{(l-1)}[s]\right) \\
    Q^{l}(t) &=W_Q^{l}\left(H^{(l-1)}[t]\right) \\
    \text{ATT-head}^{i}(s,t) &=\mathop{Softmax}\limits_{\forall s \in N(t)} \left( \frac{K^{l}(s) \cdot  Q^{l}(t)^{T}}{\sqrt{d}}\right) 
    \end{aligned}
\end{equation}
where $W_K^{l} \in \mathbb{R}^{d \times d}$ and $W_Q^{l} \in \mathbb{R}^{d \times d}$ are trainable parameters, mapping the target and source node into Query and Key vectors. The attention value of the edge $(s - t)$, $ATT-head^{i}(s,t)$, is their dot product. Then the information from node $s$ to node $t$ via edge $(s - t)$ becomes:
\begin{equation}
\nonumber
    \begin{aligned}
    V^{l}(s) &=W_V^{l}\left(H^{(l-1)}[s]\right)\\
    \text{Info}^{l}(s,t) &= W^{info} \left(\|_{i \in[1, h]}   \text{ATT-head}^{i}(s,t)\cdot V^{l}(s)\right)  
    \end{aligned}
\end{equation}
$W_V^{l} \in \mathbb{R}^{d \times d}$ is a trainable parameter which maps the source node into a Value vector, and the trainable parameter $W^{info} \in \mathbb{R}^{d \times hd}$ is to aggregate information from $h$ heads. After receiving aggregated information from all neighbours, we concatenate information from node $t$ itself and neighbours, where a trainable parameter is $W_A^{(l)} \in \mathbb{R}^{d \times 2d}$. Note that $\oplus$ means the concatenate operation. Following prior studies, we finally add a residual layer to get the embedding of target node $t$ at the $l$-th iteration as $H^{(l)}[t]$:
\begin{equation}
    \nonumber
    \begin{aligned}
    H^{(l)}[t] &= W_A^l\left(\left[H^{(l-1)}[t] \oplus\sum\limits_{\forall s \in N(t)}\text{Info}(s,t)\right]\right)+H^{(l-1)}[t]
    \end{aligned}
\end{equation}

\subsection{Incremental representation learning}
\label{sec:increment-updating}
As mentioned before, our VC network changes over time. For the newly added nodes along with all existing nodes in the time period $t$, their representations can be learned by running GST on this new network. However, this unsupervised GST without any supervision might not produce optimal embeddings, which can lead to poor performance for subsequent applications. In this study, we develop an incremental representation learning with fine-tuning, which we believe is effective for downstream success prediction. Furthermore, such an incremental strategy learns all representations across time periods into the same dimensional space, which makes these embeddings comparable.
 
We formulate this incremental representation learning problem as a supervised link prediction / node classification task as follows. (i) We use GST (denoted GST-1) to obtain representations for these newly added nodes in $G^t$ and update representations for nodes existing in $G^{t-1}$: $R^t = \text{GST-1}(E^{t-1})$. (ii) These learned embeddings $R^t$ might not be optimal, especially with a few layer aggregation from neighbors for these newly added nodes. To further optimize all embeddings, we intentionally design an optimization procedure where we use all embeddings to perform link prediction, as well as the node classification. For the link prediction, we form positive instances in the training set using all newly added links while randomly pick the same amount of other non-existing links in $G^t$ as negative instances. For the node classification problem, we have two possible labels for each node: start-up or person. The link prediction and node classification problem is formed as:
\begin{equation}
    \nonumber
    \begin{aligned}
    Dec^{LP}(u,v) &= \sigma\left(\sigma(W^LR^t[u]+b^{LP}) \cdot \sigma(W^LR^{t-1}[v]+b^{LP})\right)\\
    Dec^{NC}(v) &= \sigma(W^CR^t[v]+b^{NC})
    \end{aligned}
\end{equation}
where $R^t[*]$ is the representation of node $*$, $Dec^{LP}(u,v)$ is the link prediction decoder to predict whether a link between $u$ and $v$ is formed and $Dec^{NC}$ is the node classification decoder to identify the type of a node. The overall loss function is defined:
\begin{equation}
    \nonumber
    \begin{aligned}
    L &= \beta * L^{LP}+(1-\beta) * L^{NC}
    \end{aligned}
\end{equation}
where $L^{LP}$ and $L^{NC}$ are the binary cross entropy for the link prediction and node classification, respectively, and $\beta$ is a hyper parameter to control their contribution to the overall loss. Since we use $n$-layer GST before performing the link prediction and node classification task, this indicates the embeddings of neighbors within $n$ hops of the newly added nodes will be updated, while the embeddings of the rest remain unchanged as $E^{t-1}$. Note that when $t=0$, GST is performed based upon the randomly initialized embeddings.

\subsection{Sequential learning}
\label{sec:sequential-modeling}
After we learn graph representation for every time period, we now turn to describing how we capture their dependencies to obtain a better representation of each start-up for downstream applications, i.e. success prediction in our context. For start-ups that receive their first investments at the $t$-th time period, we model all information from the past $k$ time periods to predict whether they will succeed in the $(t)$-th period. Formally, with the learned embeddings ${E^{(t-k)},E^{(t-k+1)}, ... ,E^{(t)}}$, the sequential graph representation learning is:
\begin{equation}
    \nonumber
    \begin{aligned}
    R_{Sequential}^t &= LSTM(E^{(t-k)},\cdots,E^{(t-1)},E^{(t)})
    \end{aligned}
\end{equation}
where $R_{Sequential}^t$ is the sequential representation learned from historical embeddings, i.e., the last hidden layer of the LSTM.

\subsection{Success prediction}
\label{sec:prediction}
Now we turn towards describing the final success prediction procedure. We formulate it as a binary classification problem where the input is $R_{Sequential}^t$ learned from previous steps, plus nodes' attributes (e.g., investor demographics, industry sector). Rather directly feeding them into a three-layer Multilayer Perceptron (MLP) for success prediction, we first use the GST (denoted GST-2) to fuse them and then send the new representation to the MLP. We use the activation function of tanh and sigmoid for the first two layers and the last layer in the MLP, respectively. The training set is constructed at the start-up level where some start-ups succeed after receiving their first VC investments while others fail. The overall process of start-up success prediction is sketched in Algorithm \ref{algs:updating}.
\begin{algorithm}
\caption{Start-up success prediction}
\label{algs:updating}
\KwIn{$S^t$: VC network at time period $t$ }
\hspace{24pt} \textcolor{gray}{$V^t$: Nodes newly added to the network} \\
\hspace{24pt} {\color{gray}  $V_{changed}^t$: Nodes associated to newly added edges} \\
\hspace{24pt} {\color{gray}  $V_{neighbors}^t$: Neighbors of nodes in $V^{t}$ and $V_{changed}^t$}\\
\hspace{24pt} {\color{gray}  Node attributes: $f$}\\

\KwOut{$P$: success or failure of start-ups in $S^t$}

initialize embedding for existing nodes: $\vec{E}^{t}(S^{t-1}) \stackrel{copy}{\leftarrow} \vec{E}^{t-1}(S^{t-1})$;\\
randomly initialize embedding for $V^t$;\\

\For{each epoch $\in$ $\{1, 2, \cdots, k\}$}{
    \For{each node $v \in \{V^t, V_{changed}^t, V_{neighbors}^t\}$}{
        get the representation via GST: $R^t \leftarrow \text{GST-1}(\vec{E}^{t})$;\\
        optimize $R^t$ by minimizing $L$;
    }
}

\For{every start-up $s \in V^t$}{
    $s_{sequential} \leftarrow LSTM(E^{t-k}_s,\cdots, E^{t-1}_s, E^t_s)$; \\
    $z_s \leftarrow \text{GST-2}(f_s,s_{sequential})$; \hfill {\color{gray} \# attribute fusion}\\ 
    prediction: $P \leftarrow MLP(z_s$);
}
\end{algorithm}

\vspace{-4pt}

\section{Experiments}
\label{sec:exp}
We now introduce the details of the benchmarking dataset, followed by the evaluation of our models against the state-of-the-art baselines on success prediction. Extensive studies on model ability and  interpretability are also discussed. 

\vspace{-4pt}
\subsection{Dataset} 
\label{sec: data analysis}
In this study, we use a venture capital dataset collected by PitchBook, a subsidiary of the global financial service provider Morningstar\footnote{\url{http://www.morningstar.com}} (Alternative datasets can be obtained from other providers such as Crunchbase.). This dataset covers all VC investment-level activities taking place in main venture capital markets worldwide with a wide geographic and industry span starting from year $1977$. The data includes information about each individual investment, such as investors, investment size, and development stages of the start-ups (i.e. early, later, seed, etc). Furthermore, we enhance the dataset by collecting information regarding the demographics of entrepreneurs and VC investors, such as their geographic locations, education background, and prior employment history. Our sample consists of $187,346$ rounds of investments received by $116,764$ start-ups worldwide between year $1977$ and $2019$. We examine the prior employment and investment activities by $244,267$ founders or co-founders and $62,424$ individual VC investors (i.e. general partners). Table~\ref{table:features in pitchbook} shows the major information included in the Pitchbook dataset across three major entities: person, start-up, and investment. We perform some analysis on this dataset in Section~\ref{sec: analysis of the dataset}. 

\begin{table}[h!]
\caption{Available data in PitchBook.}
\begin{tabular}{lll}
\hline
\textbf{Start-up}  & \textbf{Person}      & \textbf{Investment}     \\ \hline
Company ID        & Person ID            & Deal ID           \\
Company Name      & Person Name          & Deal Type         \\
Year Founded      & Education Degree     & Investors         \\
Location          & Graduated Institute  & Deal Date         \\
Country           & Graduated Year       & Amount            \\
Industry Category (3 levels)  & Gender              & Company ID \\\hline
\end{tabular}
\label{table:features in pitchbook}
\end{table}
\vspace{-8pt}

\subsection{Data preparation and experiment setup}
\subsubsection{Data split.} According to our definition of success (see Section~\ref{problem_definition}), we need a five-year window to know the outcome of a start-up that has just received its first round of investment. Therefore, we use the Pitchbook data before $2015$ for training and testing, which includes $39,179$ start-ups, $85,035$ people, and $122,790$ investments. As shown in Table~\ref{tab:split}, the training set includes each first round investment from 1/1/2007 to 9/30/2008 of which we know the outcome. The validation set spans from 10/1/2008 to 12/31/2013 and can be viewed as two parts. The first part is the first 3 months and we know the outcomes for all the samples during this period. The second part covers 1/1/2009 to 12/31/2013 and this observation window of less than 5 years only allows us to identify some success cases and no failure cases can be determined. The consideration is that we need both positive and negative samples for validation and the first part provides them. The second part helps us to increase the validation set as many as possible. Our test set is from 1/1/2014 to 12/31/2014.
\label{sec:Data split}

\begin{table}[]
\caption{Data split for training, validation, and test sets.}
\begin{tabular}{lll}
\hline
Dataset                                                 & Time range            & \# of samples \\ \hline
Training                                       & 01/01/2007-09/30/2008 & 3,309 (687 positive)                          \\ \hline
\multicolumn{1}{c}{\multirow{2}{*}{Validation}} & 10/01/2008-12/31/2008 & 448 (91 positive)                            \\ \cline{2-3} 
\multicolumn{1}{c}{}                                & 01/01/2009-12/31/2013 & 1,536 (1536 positive)                           \\ \hline
Test                                            & 01/01/2014-12/31/2014 & 6,259 (805 positive)                          \\ \hline
\end{tabular}
\label{tab:split}
\end{table}

\subsubsection{Configurations.} As mentioned in Section~\ref{sec:problem definition}, the bipartite network is updated at intervals and we choose 1 month as the interval. With this setting, predictions are made monthly with only prior information. To predict the outcome for a start-up that receives its first round of funding in the $t^{th}$ time period, we only utilize features and embedding from $\mathcal{G}^t$ and its node attributes.

\subsection{Evaluation metrics}
For the binary classification results (success VS failure), we compute the Average Precision at $K$ (AP@$K$) which is used in a similar evaluation task~\cite{Bonaventura-pred-2020}. Precision at $K$ (P@$K$) measures the proportion of successful start-ups in the top $K$ start-ups suggested by the model ranked by their confidence scores and AP@$K$ is computed over 12 monthly P@$K$s in a year. A high AP@$K$ value suggests the model can potentially assist investment decision-making when investors need to screen a lot of start-ups and allocate capital in a small subset, which is the common practice in the VC industry.

\vspace{-8pt}
\subsection{Baselines}

\label{sec: embedding exp}

The models to be evaluated are grouped into two categories, namely static models and dynamic models. The former includes a random walk based embedding method (\textbf{node2vec}), a widely used graph convolutional neural network model (\textbf{GCN}), an attention based graph neural network (\textbf{GATs}), and a Gradient Boosting Decision Trees (\textbf{GBDT}) model as a traditional machine learning approach. Our method, as well as a temporal graph convolutional network model (\textbf{EvolveGCN}), falls into the latter category. As a comparison, we construct a baseline (\textbf{Human Investors}) that represents real investors' performance, calculated from our dataset. We describe the 6 baselines below.


\begin{enumerate}

\item \textbf{Human Investors}: The performance of real VC investors. We identify the start-ups in our test set that are actually selected and invested by the investors with second rounds of funding and measure their ratio of success. The ratio is 10.7\% in our test set.

\item \textbf{GBDT}: Gradient Boosting Decision Trees model uses boosting to ensemble decision trees to achieve a strong learner and it performs well in a similar start-up prediction task~\cite{sharchilev2018web}. We use XGBoost~\cite{chen2016xgboost}, an effective and efficient implementation of GBDT. Its feature vector is engineered to include both network and attribute information.

\item \textbf{node2vec}~\cite{grover2016node2vec}: A classic embedding method based on second order random walks sampling to learn node representations. This model has been widely used in applications such as node classification, clustering, and link prediction.

\item \textbf{GCN}~\cite{kipf2016semi}: A static graph neural network model, obtaining state-of-the-art performance on many graph datasets. It uses a graph convolutional layer to aggregate information from graph.

\item \textbf{GATs}~\cite{velivckovic2017graph}: Another static graph neural network model, in which the attention mechanism is used to better weight and aggregate information from neighbors. It overcomes the over smoothing problem that GCN may encounter.



\item \textbf{EvolveGCN}~\cite{pareja2020evolvegcn}: A dynamic graph neural network model, adapting the graph convolutional network model along with the temporal dimension to capture graph evolving. It often has a better performance on datasets that can be modeled with evolving graphs.



\end{enumerate}

All models are trained, validated, and tested using the data split described in Section~\ref{sec:Data split}. For each baseline, we train and validate a 3-layer MLP as described in Section~\ref{sec:prediction} using feature vectors obtained before the starting date of the test set. We then feed the feature vectors from the test set into the MLP to generate predictions. The feature vector for each node is a concatenation of its one-hot attributes and its embedding. For a person node, the attributes include gender and educational degree. A start-up node's attributes include industry sector (41 second-tier labels in total, including Software, IT Services, etc.), type of its first investment received (angel, seed, etc.), location (latitude and longitude), investment amount, and the length since its foundation. The resulting vectors have 31 and 58 dimensions for person and start-up nodes, respectively. To make a fair comparison, we refer to factors discussed in Section~\ref{sec:vc investment} and~\ref{sec: complex network} and construct feature vectors for GBDT, including start-up location, age, industry, managing members’ educational degrees (aggregated sum across different degree types), investors' educational degrees, start-up's closeness centrality, investors’ and members’ average closeness centrality, and investment amount. The configuration of models is detailed in Appendix Section~\ref{sec: exp setting}.

To evaluate each model, we pick the top predictions ranked by the output confidence values and compute AP@$K$ scores with $K$ being 10, 20, and 50, respectively. For the Human Investors baseline, its AP@$K$ scores are all set to its success ratio, $10.7\%$.



\vspace{-12pt}

\subsection{Results}

Table~\ref{tab: embedding result} reports the performance comparison among different methods. GBDT, node2vec, and GCN are all comparable to Human Investors, while the other three methods demonstrate more significant overall improvement over Human Investors. Among all methods, our model achieves the highest AP@$K$ scores, up to $1.94$ times over the Human Investors regarding AP@$10$ (i.e., $20.8\%$).

Here we examine the results in details. Though generally comparable, node2vec and GCN show an edge over GBDT, especially when $K$ is smaller (i.e. smaller numbers of start-ups are picked). We speculate that the structural information captured by node2vec may contribute more to the success prediction, as compared to limited structural information covered by manually constructed network features in GBDT, such as centralities. Furthermore, the semi-supervised nature of node2vec may have helped its $2.1\%$ improvement in AP@$20$ over the unsupervised GCN. Note that hyper-parameters involved in node2vec are chosen based on a common mechanism grid search. Both GBDT and node2vec treat neighbors equally important which is not true in practice. As observed in the results, the attention mechanism can improve the performance -- GATs achieves the highest AP@$10$ (15.0\%) and AP@$20$ (12.1\%) scores among all static methods.

Since our VC network dynamically changes, to demonstrate the superiority of our model, we also compare with a cutting-edge dynamic representation learning method EvolveGCN. As shown in Table~\ref{tab: embedding result}, our model performs better than EvolveGCN by large margins. The improvement increases as the $K$ value decreases ($16.1\%$ improvement for AP@$50$ and $24.6\%$ for AP@$10$, relatively). This indicates that our model yields more accurate predictions when the confidence value gets higher. A higher AP@$K$ value suggests our model can better assist decision-making when fund managers need to allocate capital among a selected number of start-ups, which is consistent with the common practice in the current VC industry. A plausible explanation of why our model outperforms EvolveGCN lies in three aspects, which are also the advantages of our model over EvolveGCN. (i) At each time period, our model refines the learned representation from GST through a supervised fine-tuning: link prediction and node classification; (ii) Our model incrementally updates representations for newly added nodes and their associated nodes, which can ensure the learned representations across time periods in the same dimensional space and thus they are comparable. In addition, we have a sequential learning using recurrent neural networks to further optimize the dependencies among these representations. EvolveGCN simply uses hidden states to record historical information which may not be sufficient; (iii) Unlike EvolveGCN that updates parameters (e.g., weights and biases) in every layer across time steps, our model enhances the node-level representations. We believe this is more effective because these representations are closer to the downstream applications, where together with other data (e.g., attributes) they are fused and fed to an MLP for success prediction.  

\begin{table}[]
\caption{Performance comparison among different methods.}
\vspace{-5pt}
\begin{tabular}{lllll}
\hline
Method             & AP@10  & AP@20 & AP@50 \\ \hline
Human Investors    & 10.7\%  & 10.7\%  & 10.7\%  \\
GBDT               & 11.7\%  & 10.8\%  & 10.3\%  \\
node2vec           & 12.5\%  & 12.1\%  & 11.3\%  \\
GCN                & 12.5\%  & 10.0\%  & 10.5\%  \\
GATs                & 15.0\%  & 12.1\%  & 9.8\%  \\
EvolveGCN          &  16.7\%  & 12.9\% & 11.8\%  \\
\textbf{Our model} & \textbf{20.8\%}  & \textbf{15.1\%}  & \textbf{13.7\%}  \\ \hline
\end{tabular}
\label{tab: embedding result}
\vspace{-10pt}
\end{table}

\noindent\textbf{Industry-level analysis.} As we illustrated in our related work, VC investment success prediction is likely to be heterogeneous, indicating that a model might yield different performance for different industry sectors. To understand this, we conduct an industry-level analysis by breaking down the AP@$10$ prediction results from the 120 selected start-ups (10 start-ups per month for 12 months). From Table~\ref{tab:Industry and LCC}, we can clearly see that our model performs well for start-ups in IT, healthcare, and B2C industries, and even triples the performance of Human Investors in healthcare. By further investigating the network, we find that nodes in the VC networks associated with these three industries are mostly falling into the Largest Connected Component (LCC). The other 4 industries, on the other hand, are not well-connected measured by the percentage of nodes in LCC and 3 of them are much smaller in sizes. This is consistent with our analysis of the dataset regarding LCC in Appendix Section~\ref{sec: analysis of the dataset} and it suggests that network representation learning-based methods lead to better performance for relatively dense networks in the context of VC investment success prediction.


As a side note, we also perform an analysis on hyper-parameter (i.e. GST layers, embedding dimension, and loss function weight $\beta$) sensitivity, described in Appendix Section~\ref{sec: hyper-parameter}.

	
\begin{table}[]
\caption{Performance comparison for different industry sectors. (M\&R: Materials and Resources, FS: Financial Services and HI: Human Investors.)}
\vspace{-8pt}
\begin{tabular}{lllll}
\hline
Industry   & Precision & \% in LCC & HI & Selected/total \\ \hline
IT         & 23.9\%      & 73.9\%      & 13.9\%   & 67/3061=2.2\%    \\
Healthcare & 36.4\%      & 68.9\%      & 10.8\%   & 11/779=1.4\%    \\
B2C        & 26.3\%      & 67.8\%      & 9.3\%   & 19/1387=1.4\%    \\
FS         & 0.0\%      & 63.3\%      & 10.7\%   & 1/150=0.7\%     \\
B2B        & 0.0\%      & 61.0\%      & 9.8\%   & 20/818=2.4\%    \\
M\&R       & 0.0\%      & 54.2\%      & 8.4\%   & 1/83=1.2\%    \\
Energy     & 0.0\%      & 51.9\%      & 7.8\%   & 1/129=0.8\%     \\ \hline
\end{tabular}
\label{tab:Industry and LCC}
\vspace{-10pt}
\end{table}


\vspace{-4pt}
\section{Discussion}
\label{sec:discussion}
In this section, we further explore the results from the experiments and discuss implications for VC investment practice, policy making, and relevant academic research. We do this by examining the characteristics of important people associated with positive predictions.

We use self-attention of GST-2 (described in Section~\ref{sec:prediction}) to identify important people who influence the positive predictions and compare them with the average from the test set. For each start-up in the test set that connects more than one person in the network, we focus on the affiliated person with the highest attention value. 890 of the 7,268 start-ups in the test set are predicted to be successful (positive predictions) by our method. For these 890 start-ups, we get 871 people in total (One person can be identified by multiple start-ups.) and notice the following salient patterns.


\begin{table}[]
\caption{Gender and degree distributions for people associated with positive success predictions by our method, and the ratios for people in the whole test set.} 
\vspace{-8pt}
\begin{tabular}{llll}
\hline
\multicolumn{2}{l}{}             & \% in test set & \% in selected people \\ \hline
\multirow{4}{*}{Degree} & Bachelor     & 19.4\%              & 29.0\%                 \\
                        & Master       & 28.3\%              & 31.0\%                 \\
                        & Doctor       & 11.2\%              & 14.1\%                 \\
                        & Other        & 41.1\%              & 25.8\%                  \\ \hline
\multirow{2}{*}{Gender} & Male   & 92.0\%              & 89.9\%                 \\
                        & Female & 8.0\%               & 10.1\%                 \\ \hline
\end{tabular}
\label{tab:Degree of selected people}
\vspace{-16pt}
\end{table}

First, people associated with positive predictions have more advanced education backgrounds. As shown in Table~\ref{tab:Degree of selected people}, for all people in the test set, 58.9\% have Bachelor degrees or above while this ratio goes up to 74.2\% for the selected people. The ratio for Doctor degrees also arises from 11.2\% to 14.1\%. This suggests the value of academic training to VC investment outcomes, though we cannot separate a positive effect of acquiring an advanced degree from the selection effect of education on the intrinsic capability of people.

In Table~\ref{tab:Degree of selected people}, we also note the higher percentage of female in the selected sample associated with positive predictions (i.e. $10.1\%$), compared to a percentage of female percentage of $8.0\%$ in the entire sample. The lack of gender diversity in VC industry has been long-standing, as male investors or managing members far outnumber their female counterparts ~\cite{gompers2017diversity,gompers2014gender}. Our finding suggests a positive association between increased gender diversity and success in VC investments, consistent with prior finding in ~\cite{gompers2017diversity}. Apart from a direct effect of gender diversity on performance, we contend such positive relationship is also likely driven by selection of very capable female investors or female managing members that have managed to thrive in this male-dominated field. Existing research shows that male investors hold bias against female managing members when making investment decisions ~\cite{ewens2020early}. Furthermore, only about 10\% of new hires in the venture capital industry are women, and approximately 75\% of venture capital firms have never had a senior investment professional who is a woman ~\cite{gompers2014gender}. Therefore, the very few numbers of female investors or female-led start-ups that received funding in the data are likely endowed with superb capabilities, and thus, present higher probability of success compared to their male counterparts.          

Last, we find people associated with positive predictions  have higher degree centrality in the network. Specifically, the average degree centrality of these people is 3.85, compared to an average degree centrality of 3.20 for all the people in the test set, which is consistent with the literature acknowledging the importance of networking to success of entrepreneurship~\cite{HOCHBERG-whom-2007,hoang2003network}.

\vspace{-8pt}
\section{Conclusion}

\label{sec:conclusion}

In this paper, we present an incremental graph representation learning to address the challenging task of predicting whether an early-stage start-up would eventually succeed (IPO or acquired), which is crucial for venture capitals and policy makers. Our model learns better node-level representation in each time period using self-attention and fine-tuning via supervised link prediction and node classification, as well as modeling sequential dependencies of node representations across time periods. Our model also integrates the network structural information and the attributes of investors, start-ups, and the managing members of the start-ups, archiving state-of-the-art overall performance as compared to several baselines. We also show that our method is particularly effective for those nodes located in the Largest Connected Component. By further investigating the results, we discover insights on predictive factors such as gender, education, and networking.

For future work, there are many directions. First, we plan to incorporate crowd-sourced textual data for depicting the start-ups' competitions and popularity. Second, we intend to collect more data to track the new start-ups that have not yet received any types of VC investments and predict their outcomes.



\vspace{-8pt}
\bibliographystyle{ACM-Reference-Format}
\bibliography{sample-base}

\clearpage

\appendix

\huge \textbf{APPENDIX}

\normalsize


\section{Data Analysis}
\label{sec: analysis of the dataset}
\begin{figure}[h!]
\centering
\includegraphics[width=8cm]{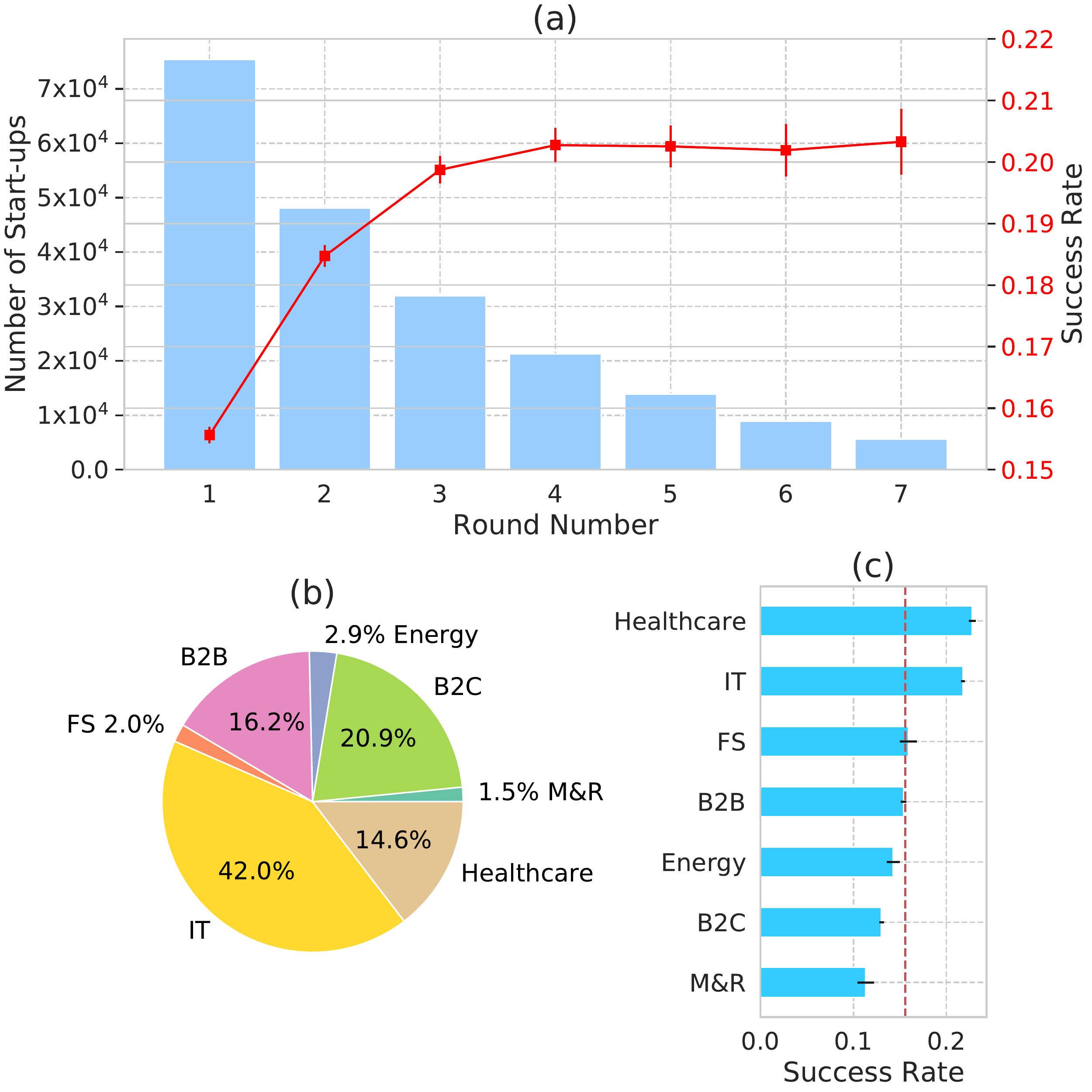}

\caption{ a) The numbers of start-ups at different rounds of funding and their corresponding success rates with standard error. b) The distribution of start-ups across different industry sectors, with FS being Financial Services and M\&R being Materials and Resources. c) The success rates of different industry sectors with standard error. The average success rate across all sectors is marked by the red dotted line.}
\label{fig:success rate}
\end{figure}

We perform some analysis on this dataset. Figure ~\ref{fig:success rate}a visualizes the numbers of start-ups for each round and the corresponding success rates. It indicates that foreseeing the outcomes for the first round start-ups is the most difficult with a success rate of $15.71\%$ and the difficulty decreases when the start-ups enter later rounds.

When we look into different industry categories (first-tier categories, 7 in total), we see that $93.6\%$ of the start-ups fall into IT, B2C, B2B, or healthcare (Figure ~\ref{fig:success rate}b) and start-ups in healthcare and IT achieve higher success rates (Figure~\ref{fig:success rate}c).

\begin{figure}[h!]
\centering
\includegraphics[width=8.5cm]{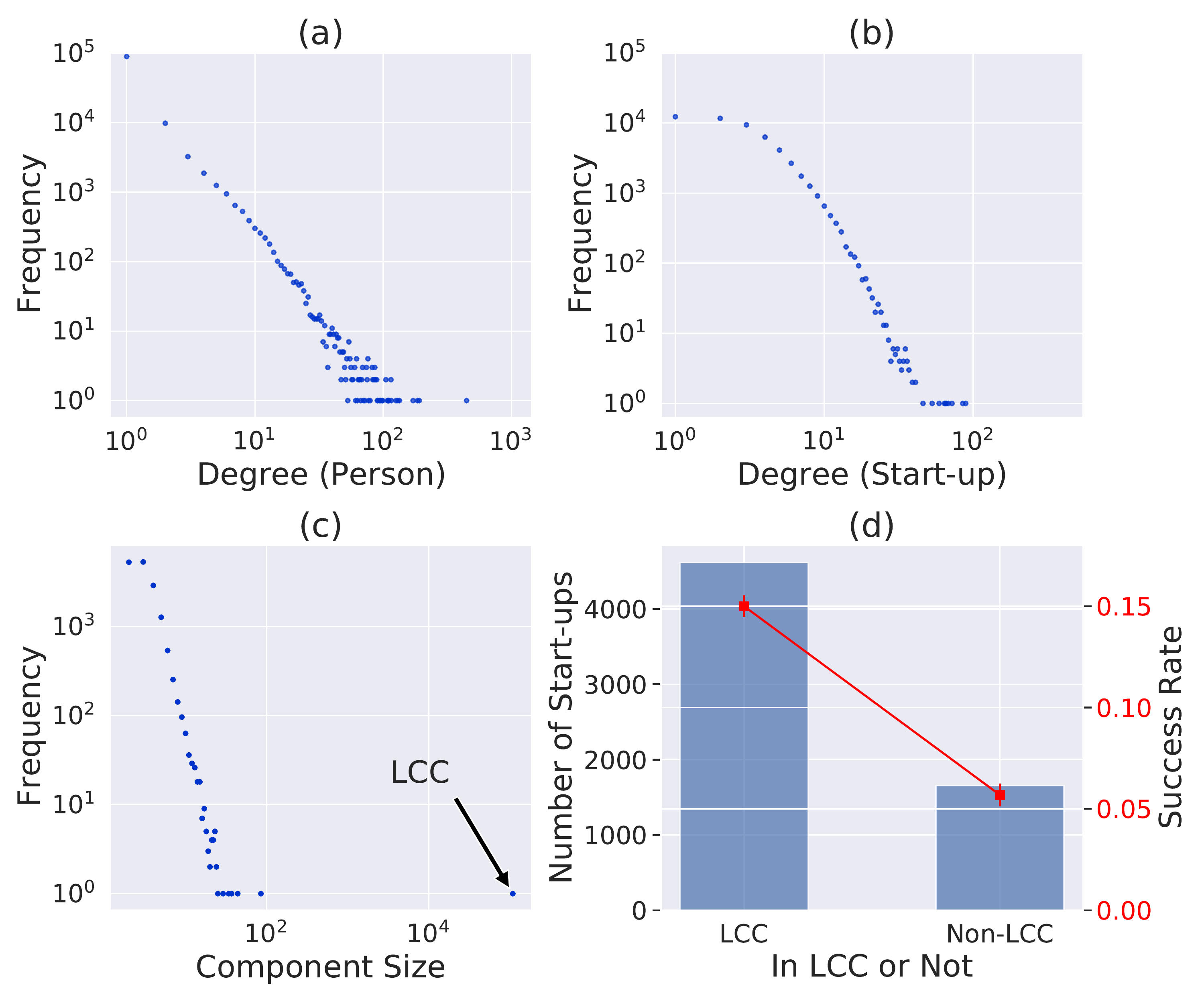}
\caption{a) People's degree distribution in the VC Network. b) Start-ups' degree distribution.  
c) Component size distribution. d) Numbers of start-ups in the Largest Connected Component versus other components, with success rates.
}
\label{fig:graph statistic}
\end{figure}

From a network perspective, the log-log degree distributions for person (Figure ~\ref{fig:graph statistic}a) and start-up (Figure ~\ref{fig:graph statistic}b) indicate strong power-laws. When we examine the sizes of the components in our VC network, we discover that the Largest Connected Component (LCC) has $66.56\%$ ($108,573$) of the nodes while the Second Largest Connected Component only takes up $0.05\%$ ($85$) in a 
network snapshot on 12/31/2014, as shown in Figure ~\ref{fig:graph statistic}c. Following this huge polarization, we further calculate the average success rate for first-round start-ups in the Largest Connected Component versus other first-round start-ups in $2014$. The former ($15.00\%$) is statistically significantly higher than the latter ($5.69\%$). This demonstrates that network structural information is indicative for start-ups' success in our dataset, which is consistent with prior studies.

\section{Model Configuration}
\label{sec: exp setting}
For the hyper-parameter we used in experiments, the embedding size is set to 64, the number of heads for multi-head attention is 8, the GNN-based method uses a 3-layer GNN, and the $\beta$ used in the loss function is set to 0.5. For the baseline methods, we use the node classification task to train the embeddings, and we train the node2vec with the walk length and number of walks being 10 and 80, respectively.

For dynamic graph neural network methods, we use data before 1/1/2007 to build a network with random embedding initialization and the data from 1/1/2007 to 1/1/2015 to update the embeddings. For static methods, we use data before 1/1/2015 to build and embed the network and each sample's embedding is created all at once, regardless of their chronological order.
 
\section{Hyper-parameter Sensitivity}
\label{sec: hyper-parameter}
\begin{figure}[h!]
\centering
\includegraphics[width=9cm]{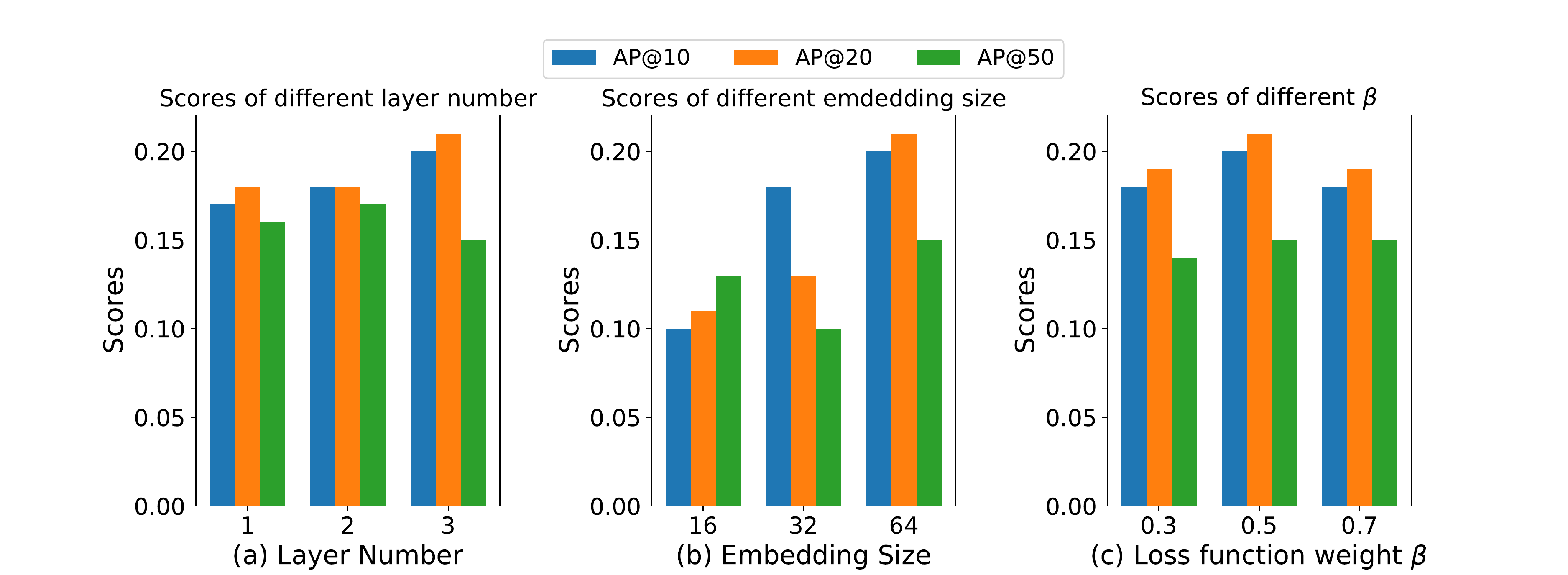}
\caption{Results of our methods with different hyper-parameters}
\label{fig: Hyperpara}
\end{figure}

In this section, we examine the influence of key hyper-parameters on the performance of our methods. When studying one parameter, others are fixed. Other settings are kept the same as described in Section~\ref{sec: embedding exp}. Figure~\ref{fig: Hyperpara} illustrates the AP@$K$ metrics for different hyper-parameters.

\begin{itemize}

    \item \textbf{Layers of GST}: When we vary GST's number of layers from 1 to 3, the overall performance increases as GST becomes deeper, suggesting more useful information may be encoded from the VC network.
    
    \item \textbf{Dimension of embedding}: We increase the dimension of embedding $d$ from 16 to 64 and it suggests that a larger $d$ can boost the overall performance of the start-up success prediction. However, due to the heavy consumption of GPU memory, we cannot increase it further.
    
    \item \textbf{Loss function weight $\beta$}: When the loss function weight $\beta$ is increased from 0.3 to 0.7, we are re-balancing the weights for the two loss functions. Figure~\ref{fig: Hyperpara} shows that the best overall performance is achieved when they are equally weighted. Without any loss function being dominant, multiple loss functions may contribute more to the overall performance improvement.

\end{itemize}
\end{document}